\documentclass{fairmeta}
\usepackage{amsmath}
\usepackage{enumerate} 
\usepackage{bm}
\usepackage{algorithm}
\usepackage{floatflt}
\usepackage{algpseudocode}
\usepackage{amsfonts}
\usepackage{amsthm}
\usepackage{newtxtt}
\usepackage{algorithm}
\usepackage{colortbl} 
\usepackage{cleveref}
\usepackage{diagbox} 
\usepackage[utf8]{inputenc}
\usepackage{textgreek}
\usepackage{colortbl}
\usepackage{nicematrix}
\usepackage{makecell}
\usepackage{float}
\usepackage{arydshln}
\usepackage[frozencache,cachedir=.]{minted}
\usepackage{caption}
\usepackage{subcaption}
\usepackage{tcolorbox}
\usepackage{amssymb}
\usepackage{xspace}
\usepackage{wrapfig}
\usepackage{adjustbox}
\usepackage{tabularx}
\usepackage{booktabs}
\usepackage{mathtools}
\usepackage{wrapfig}
\usepackage{amssymb}
\usepackage{graphicx}

\usepackage{silence}
\makeatletter
\patchcmd{\wrong@fontshape}{\@gobbletwo}{}{}{}
\makeatother
\definecolor{upColor}{RGB}{17,138,21}
\definecolor{downColor}{RGB}{174,36,67}

\newtheorem{theorem}{Theorem}[]

\newtheorem{remark1}[theorem]{Remark}

\title{Point2Radio: A Foundation Model for Cross-Scene Radio Fields from Material-Aware Point Clouds}
\author[1]{{Chaozheng Wen}\textsuperscript{$\dagger$}}
\author[1]{{Chenghong Bian}\textsuperscript{$\dagger$}}
\author[1]{Hongze Chen}
\author[1]{Jun Zhang\textsuperscript{*}}
\affiliation[1]{Hong Kong University of Science and Technology}

\contribution{\textsuperscript{$\dagger$}Equal Contributions$\quad$\textsuperscript{*}Corresponding Authors}
\code{\url{https://github.com/wenchaozheng/Point2Radio}}
\metadata[Correspondence to]{Jun Zhang (\email{eejzhang@ust.hk})}

\begin{document}

\abstract{
High-fidelity radio fields are typically simulated for every
scene--transmitter configuration or fitted separately to each scene, failing to exploit propagation structures shared across environments. We present
\textbf{Point2Radio}, a foundation model that learns a transferable propagation prior from
multiple environments. Given a material-aware point cloud and a transmitter (TX)
setting, a common encoder produces a TX-conditioned scene
representation that can be queried at arbitrary receiver (RX) locations.
Task-specific query decoders map this representation to different radio
quantities, e.g., three-dimensional (3D) path-gain (PG)
fields and power angular spectra (PAS). At inference for a new scene, the model uses only a
material-aware point cloud and transceiver queries, running in milliseconds
on a single GPU without meshes or explicit path tracing. We evaluate PG prediction on a scene-disjoint split of a
337-scene corpus containing 86{,}272 TX-conditioned fields. 
 Point2Radio\ achieves 0.871\,dB mean absolute error (MAE), reducing error by 76.7\% relative
to a same-split UNet-style baseline. The same encoder also supports PAS
prediction via a task-specific decoder. Experiments further show that light target-scene fine-tuning
improves adaptation to a specific environment.

}

\maketitle

\section{Introduction}

Modeling radio fields in 3D environments is increasingly
important for low-altitude aerial networks, embodied AI, connected autonomous
vehicles, and multi-robot coordination \cite{10430216}. It can guide
coverage-aware planning and reliable multi-agent communication, while RF
observations offer embodied agents an additional sensing modality \cite{guo2025omnivla}. These
applications all benefit from knowing how a transmitted signal arrives across a scene. Radio field prediction is a long-standing problem, yet accurate and transferable solutions remain scarce. Radio waves undergo reflection,
transmission, scattering, and diffraction on complex surfaces and materials.
These interactions rarely admit accurate closed-form solutions, so high-fidelity
fields are usually obtained only through expensive numerical simulation or dense
on-site measurement. 

\begin{figure*}[t]
\centering
\includegraphics[width=0.98\textwidth]{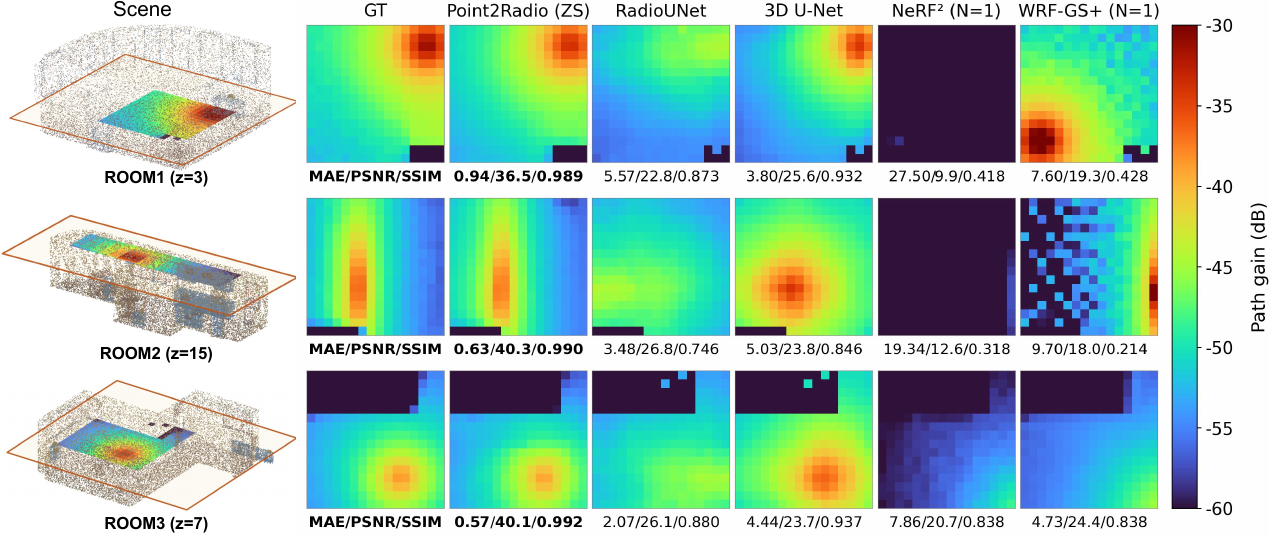}
\caption{Visualization of cross-scene generalization on three completely unseen
indoor rooms. Left: 3D  point cloud with
the selected height slice marked by the cutting plane. Remaining columns:
ground-truth (GT) PG map and predictions from zero-shot (ZS) Point2Radio, RadioUNet,
3D~U-Net, NeRF$^2$, and WRF-GS+. Slice-wise MAE$\downarrow$ / peak
signal-to-noise ratio (PSNR)$\uparrow$ / structural similarity
(SSIM)$\uparrow$ are reported under each map. Since
NeRF$^2$ and WRF-GS+ are per-scene methods without cross-scene
transfer, we train each with a single randomly chosen TX sample (N=1) for a
fair low-data comparison against ZS baselines. }
\label{fig:vis}
\end{figure*}

Recent progress in AI and foundation models has transformed vision \cite{carion2025sam} and language \cite{achiam2023gpt},
but has transferred less cleanly to radio-field modeling. Radio labels are
often scarce, costly to collect, and inconsistent across devices, frequencies,
and sites. As a result, learning systems in this area still struggle to acquire
broad, reusable propagation knowledge.

Existing approaches attack the problem from several directions. Classical empirical models \cite{sarkar2003survey} are inexpensive but cannot resolve detailed spatial
variation. Deterministic ray tracing instead constructs propagation paths and
applies parameterized interaction models. Increasing the ray density, path
depth, and supported mechanisms improves fidelity, but computational cost grows
rapidly, and the search must be truncated and repeated for every new scene and
TX configuration \cite{hoydis2023sionna}. This fidelity--computation trade-off
motivates learning recurring propagation patterns from expensive offline
simulation and reusing that experience across configurations, rather than
recomputing paths at inference.

Learning-based methods currently occupy two main operating points. Amortized
cross-scene predictors such as RadioUNet \cite{RadioUNet} are feedforward and transferable, but
typically map two-dimensional (2D) layouts to fixed path-loss rasters and thus omit
the 3D geometry and material interactions that govern multipath
propagation. Scene-specific neural RF representations take the
opposite trade-off: methods such as NeRF\(^{2}\) \cite{zhao2023nerf2} and WRF-GS \cite{wen2025wrf} recover detailed
fields within one environment, but require target-scene measurements or
optimization and do not reuse a single model across scenes. Existing work thus tends either to amortize across scenes
while compressing geometry into 2D, or to retain 3D fidelity by
refitting each environment.

These limitations invite a natural question: can a model learn transferable
priors about how radio signals interact with 3D scenes, rather
than treating each environment as a fresh computation or fit? If such priors
are captured once, cross-scene reuse becomes possible, and different decoders
can be trained on the same representation for different downstream radio
quantities. We therefore cast radio fields prediction in the usual
foundation model form of a shared backbone with task-specific heads:
\begin{equation}
H = E(R,\mathbf{t}),
\qquad
\hat{y} = D(H,\mathbf{q}).
\label{eq:foundation-formulation}
\end{equation}
Here \(R\) denotes a structured 3D scene representation that carries geometry
and electromagnetic material information, \(E\) encodes \(R\) together with the
TX setting \(\mathbf{t}\) into a scene code \(H\), and a task-specific decoder
\(D\) answers query \(\mathbf{q}\).  In principle, any radio quantity queried at
an RX from \((H,\mathbf{q})\) fits this interface. 

We propose \textbf{Point2Radio} as one realization of this paradigm. We first
represent each scene as a material-aware point cloud. Structured hierarchical
tokenization then aggregates scene points into a compact token set. These
tokens are fed with the TX into Cross--Self--Cross (CSC) attention, which
yields TX-conditioned latent codes. At inference, each query gathers nearby
codes by $k$NN and a task-specific head predicts the radio quantity. The shared encoder transfers ZS to unseen scenes, and can be further
adapted by light decoder fine-tuning when target scene labels are available.
This paper validates the design on PG prediction for dense spatial
fields (Figure~\ref{fig:vis}) and PAS prediction for directional spectra (Figure~\ref{fig:aoa_vis}).

Our main contributions are:
\begin{enumerate}
    \item We present Point2Radio, a foundation model for cross-scene radio fields
    that predicts RX-centric quantities from a material-aware point cloud
    and a TX setting. Hierarchical tokenization and CSC attention encode
    TX--scene interactions into transferable latent codes.

    \item We show that a PG-pretrained encoder can be reused across tasks by
    attaching task-specific query decoders while keeping the backbone frozen,
    enabling fast transfer from PG to PAS. Light residual adaptation
    further improves performance on a target scene.

    \item We introduce \textbf{PRISM}, a
    \textbf{P}rocedural \textbf{R}adio--\textbf{I}mage \textbf{S}cene
    \textbf{M}ultimodality dataset, which provides aligned wireless and vision
    annotations on procedurally generated indoor scenes. This paper uses the
    wireless splits PRISM-PG, with dense PG volumes for 337 scenes,
    and PRISM-PAS for directional experiments.

    \item Extensive experiments on PG and PAS prediction show that
    Point2Radio\ is both accurate and efficient, enabling millisecond single-GPU inference with simple inputs and low compute cost.
\end{enumerate}

\section{Related Work}

\subsection{Per-scene radio field prediction.}
Radio propagation depends strongly on the surrounding geometry and materials,
which has motivated transferring neural scene representations from computer
vision to per-scene radio field modeling.
Neural radiance fields (NeRF) \cite{mildenhall2021nerf} represent a scene as a
continuous field optimized from sparse observations. This paradigm has been
adapted to wireless channel prediction within a site
\cite{3692070.3693415, 11571284}. More recently, 3D Gaussian splatting (3DGS)
\cite{kerbl3Dgaussians} replaces implicit MLPs with explicit Gaussian
primitives for faster high-quality reconstruction. Subsequent radio works
adopt 3DGS for spatial spectrum reconstruction, with physical propagation priors or visual priors \cite{11258087,11355734}. Closely related
extensions keep a scene-level Gaussian representation while broadening the
radio target, including cross-frequency radiation fields with shared geometry
and frequency-adaptive RF attributes \cite{wang2026xfreq}, and geometry-conditioned
delay--beam priors for high-mobility channel estimation \cite{zhang2026geogs}.
A common bottleneck is that many of these reconstructions specialize to a
fixed TX or a fixed RX setting. 

Subsequent methods improve TX--RX flexibility inside a reconstructed scene.
RFCanvas \cite{10.1145/3666025.3699351} adapts the model with visual priors and few-shot radio measurements. URF-GS  \cite{wen2026bridging} uses physics-informed inverse rendering
to separate emission from environment properties and synthesize new TX--RX
pairs. RayProNet learns a neural point-field surrogate
that can be queried at new TX/RX locations after scene-specific training
\cite{10684152}. Even with such within-site flexibility, each new environment
still requires its own optimization. In contrast, Point2Radio\ learns a
transferable scene representation for scene-disjoint prediction, without
target-scene fitting.

\subsection{Cross-scene radio field prediction.}
Differentiable ray-tracing methods take another path to new
environments. Once a scene is available, they \cite{orekondy2023winert,Chen_2025_CVPR} can synthesize
channels for new TX--RX links through an explicit 3D propagation loop. Radio digital twins \cite{11115919} similarly couple mapped geometry
with learned electromagnetic materials in a differentiable tracer. These approaches have a degree of scene-level generalization ability, but the
transfer is geometry-conditioned. High-fidelity meshes and material information are often difficult to obtain at scale, and new sites commonly still need sparse measurements to calibrate materials or interaction parameters.

Cross-scene radio map networks instead pursue a feedforward manner without
specific scene fitting, predicting path-loss coverage from
TX-conditioned layout rasters \cite{10437562}. Generative models further
extend this paradigm to dynamic or multi-height radio-map tensors
\cite{11083758}. While they avoid per-scene ray tracing, their
inputs remain occupancy-style maps and their outputs are discretized rasters,
so fine-grained 3D surface geometry and materials are easy to lose.
Another line of work trains foundation models on wireless measurements. WiFo \cite{liu2025wifo}
pretrains a masked model to complete channel tensors from partial observations. MUSE-FM  \cite{zheng2026muse} learns a shared backbone for several downstream radio
tasks and conditions on a 2D top-down layout map.
These designs remain limited in capturing fine-grained 3D geometry and
materials for dense spatial field queries. Relative to this cross-scene
line, Point2Radio\ takes a material-aware point cloud rather than a layout raster,
measurement tensor, or calibrated mesh. It predicts radio quantities at dense
3D query points in one feedforward pass without online ray tracing or
target-scene radio fitting.

\section{Method}

\begin{figure*}[t]
\centering
\includegraphics[width=0.95\textwidth]{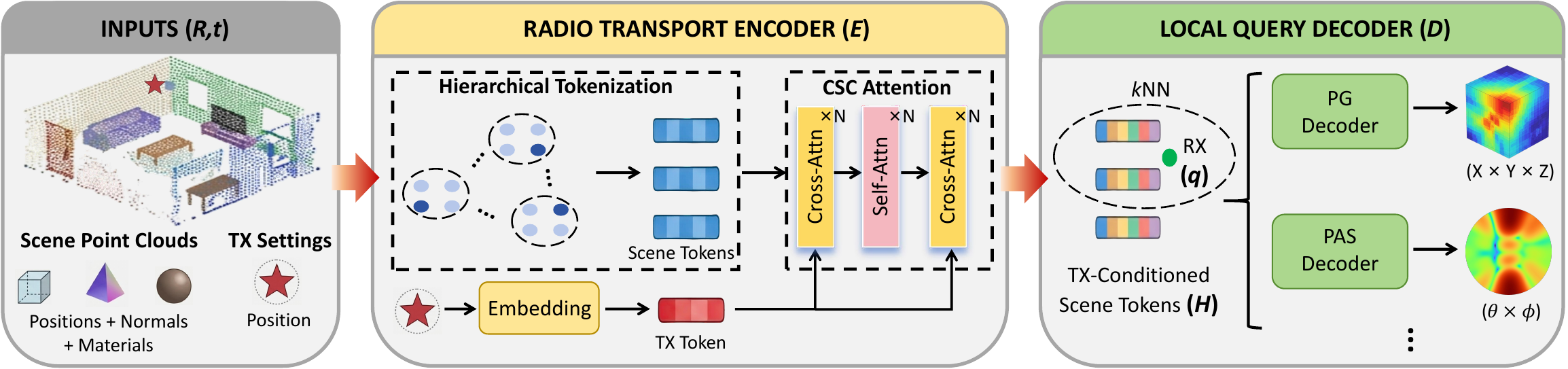}
\caption{Overview of Point2Radio. Inputs are a material-aware point cloud and a TX
setting. The encoder tokenizes the scene and applies CSC attention to form
TX-conditioned latent codes. The local query decoder aggregates nearby features
for task-specific PG or PAS prediction.}
\label{fig:pipeline}
\end{figure*}

In this section, we present Point2Radio, which realizes the encoder--decoder
process in Eq.~(\ref{eq:foundation-formulation}), as depicted in
Figure~\ref{fig:pipeline}. Given a material-aware point cloud and a TX
setting, we first construct the scene representation \(R\) by attaching
electromagnetic attributes and an explicit TX emission point. The encoder
\(E\) then hierarchically tokenizes the scene and fuses tokens with the TX
through CSC attention, producing TX-conditioned latent codes \(H\). Finally,
the decoder \(D\) applies the same local \(k\)NN query mechanism at the RX
location for both PG and PAS, and differs mainly in the task heads that map
the aggregated features to scalar PG or directional PAS.  Finally, we attach a
lightweight residual head for fine-tuning, enabling fast adaptation to a new
target scene.

\subsection{Input Representation}

The pipeline begins by constructing the scene representation for the
encoder. We represent each indoor environment as a material-aware point cloud.
Compared with structured geometry such as meshes or volumetric grids, point
clouds are more flexible and more accessible. The sampling density can be
adapted to scene complexity, enabling a balance between efficiency and
accuracy. Moreover, the absence of explicit connectivity among points
simplifies subsequent neural processing.

In practice, each scene is discretized into \(M\) surface
points. Every point is assigned a geometric and electromagnetic feature:
\begin{equation}
R=
\left\{
(\mathbf{x}_i,\mathbf{n}_i,\epsilon_i,\sigma_i,s_i,\chi_i,e_i)
\right\}_{i=1}^{M}
,
\end{equation}
where \(\mathbf{x}_i\) and \(\mathbf{n}_i\) denote the position and surface
normal. Relative permittivity \(\epsilon_i\) and conductivity \(\sigma_i\)
follow the ITU material model~\cite{itu-p2040}. For numerical stability, and to
avoid domination by large values such as metal conductivity
(\({\approx}10^{7}\,\mathrm{S/m}\)), we store \(\sigma_i\) as
\(\log_{10}\sigma_i\). A scattering coefficient \(s_i\) and a
cross-polarization discrimination coefficient \(\chi_i\) are further attached
according to the material class. To distinguish scene points from the TX, we
add a binary emission indicator that is zero on every point in \(R\) and one on
the TX. The TX setting \(\mathbf{t}\) is then formed from its location
together with this emission channel.  The pair \((R,\mathbf{t})\) is then
passed to the radio transport encoder \(E\), which
maps it to TX-conditioned latent codes \(H\).

\subsection{Radio Transport Encoder}

Given \((R,\mathbf{t})\), the encoder \(E\) produces a reusable TX-conditioned
scene code \(H\). Explicit path tracing is costly at inference and does not
learn a transferable prior. Following~\cite{10.1145/3799902.3811095},
multi-bounce transport can be viewed as repeated scene-wide mixing of an
emission under a transport operator, and attention offers a related
all-to-all aggregation as a feedforward surrogate. The same operator view
carries over to indoor radio, where a TX emission reaches RXs through
multipath interactions with material surfaces, including reflection,
diffraction, and transmission. We therefore use attention for TX-conditioned
propagation encoding rather than online path tracing. To this end, we design
\(E\) to model propagation between the TX and scene points through
hierarchical tokenization followed by CSC attention, as
shown in Figure~\ref{fig:pipeline}.

\paragraph{Hierarchical tokenization.}
Attending over every point in \(R\) is too expensive, since self-attention
scales quadratically with the number of tokens. We compress \(R\) into fewer
tokens while retaining local geometry and material cues. A shared MLP first
lifts per-point features. We then apply two downsampling stages. Each stage
selects centers by farthest-point sampling (FPS), gathers \(k\) nearest
neighbors for every center, and aggregates neighbor features with an MLP
followed by max-pooling. This yields \(N_t\) tokens
\(\mathbf{S}=\{\mathbf{s}_j\}_{j=1}^{N_t}\) of width \(d\), located at the
retained centers \(\{\mathbf{x}_j\}_{j=1}^{N_t}\). The TX setting
\(\mathbf{t}\) is encoded by a separate MLP into \(\mathbf{e}_{\mathrm{tx}}\).
We write \(\mathbf{x}_{\mathrm{tx}}\in\mathbb{R}^{3}\) for the TX location
in \(\mathbf{t}\), which later provides relative geometry for TX--scene and
TX--RX terms.

\paragraph{TX-conditioned CSC attention.}
CSC models TX--scene interactions and stores the result in the scene tokens
for later RX decoding. As shown in the Figure~\ref{fig:pipeline}, we
stack TX-to-scene cross-attention, scene self-attention, and another
TX-to-scene cross-attention, each repeated \(N\) times. The first cross-attention
brings TX information into the scene tokens, self-attention exchanges
information among tokens, and the second cross-attention updates the tokens
again with the TX. Each TX-to-scene block uses vector cross-attention with
relative geometry:
\begin{equation}
\Delta\mathbf{x}_j=\mathbf{x}_{\mathrm{tx}}-\mathbf{x}_j,\qquad
\mathbf{P}_j=\gamma(\Delta\mathbf{x}_j),
\end{equation}
\begin{equation}
\mathbf{q}_j=\mathbf{W}_q\mathbf{s}_j,\qquad
\mathbf{k}=\mathbf{W}_k\mathbf{e}_{\mathrm{tx}},\qquad
\mathbf{v}=\mathbf{W}_v\mathbf{k},
\end{equation}
\begin{equation}
\mathbf{A}_j=\mathrm{softmax}\big(\psi(\mathbf{k}-\mathbf{q}_j+\mathbf{P}_j)\big),
\end{equation}
\begin{equation}
\tilde{\mathbf{s}}_j=\mathbf{W}_o\big(\mathbf{A}_j\odot(\mathbf{v}+\mathbf{P}_j)\big).
\end{equation}
Here \(\gamma\) and \(\psi\) are small MLPs,
\(\mathbf{W}_q,\mathbf{W}_k,\mathbf{W}_v,\mathbf{W}_o\) are learned
projections, and \(\odot\) is element-wise multiplication. The output
\(H=\{\mathbf{s}_j\}_{j=1}^{N_t}\) remains anchored at
\(\{\mathbf{x}_j\}_{j=1}^{N_t}\).

\subsection{Local Query Decoder}

After CSC, each retained token stores a TX-conditioned code at a surface
location. Interpreting these tokens as virtual scatterers, we decode each RX
by aggregating the nearest codes around the query. Since CSC self-attention
has already propagated long-range scene context across tokens, this local
neighborhood contains both local geometric cues and non-local propagation
information, making local aggregation a principled readout.

\paragraph{Local query aggregation.}
Given an RX location \(\mathbf{q}\), we first build a query embedding with an
MLP over the concatenated features
\begin{equation}
\big[
\mathbf{q},\;
\mathbf{q}-\mathbf{x}_{\mathrm{tx}},\;
\|\mathbf{q}-\mathbf{x}_{\mathrm{tx}}\|
\big],
\end{equation}
which encode the RX position together with its displacement and distance to
the TX. Next, we select the \(K\) codes in \(H\) whose anchors are nearest to
\(\mathbf{q}\). Finally, the query embedding attends to these neighbor codes
with the same vector cross-attention as in CSC, using the relative
displacements from \(\mathbf{q}\) to the neighbor anchors as the positional
term. The updated embedding is the RX latent fed to the task head.

\paragraph{Task heads.}

PG and PAS share the encoder and the local readout, and differ only in the
task head applied to the RX latent. For PG, the query is an arbitrary 3D
location \(\mathbf{q}\). An MLP maps the RX latent to a scalar PG
\(\hat{y}_{\mathrm{pg}}(\mathbf{q})\). For PAS, we keep the same RX latent
and further query a direction of azimuth \(\phi\) and elevation \(\theta\).
Let \(\mathbf{u}(\phi,\theta)\) be the corresponding unit vector. We form
Fourier features
\begin{equation}
\xi(\mathbf{u})
=
\big[
\mathbf{u},\;
\{\sin(2^{k}\pi\mathbf{u}),\,\cos(2^{k}\pi\mathbf{u})\}_{k=0}^{L-1}
\big],
\end{equation}
embed \(\xi(\mathbf{u})\) with an MLP, concatenate the result with the RX
latent, and map the fused features to directional power
\(\hat{y}_{\mathrm{pas}}(\mathbf{q},\phi,\theta)\). After PG pretraining, we
freeze the spatial encoder and train only this PAS head, transferring the
learned propagation prior to directional spectra.

\paragraph{Target-scene residual adaptation.}
ZS inference already transfers a shared propagation prior to unseen rooms,
but a held-out scene can still exhibit systematic offsets from material
mismatch, unmodeled clutter, or simulator--reality gaps. When a few labeled
links are available in that target scene, we therefore keep the pretrained
encoder and task head frozen and attach a lightweight residual head that
corrects only the residual error. Let \(\hat{y}_0\) denote the frozen
prediction at a query. The adapted output is
\begin{equation}
\hat{y}=\hat{y}_0+r(\mathbf{g}),
\end{equation}
where \(r\) is a small MLP and \(\mathbf{g}\) collects TX--RX geometry used by
the residual, including the RX location, the displacement
\(\mathbf{q}-\mathbf{x}_{\mathrm{tx}}\), and the link distance. For PG, \(r\)
predicts a scalar correction in dB; for PAS, the same geometric features feed
a compact spectrum head that outputs an additive angular residual map. The
final layer of \(r\) is zero-initialized, so adaptation begins as an identity
mapping and cannot degrade the frozen prior before training. Only the residual
parameters are updated on the target-scene labels, which keeps adaptation
cheap, preserves the cross-scene representation learned by the encoder, and
applies unchanged to both PG and PAS.

\paragraph{Training objective.}
Learning proceeds in stages that mirror this encoder--decoder factorization.
We first pretrain the shared encoder and PG head with mean squared error on
normalized PG targets, so the backbone absorbs a cross-scene propagation
prior. For PAS, we then freeze the encoder and optimize only the directional
head on spectrum labels. When target-scene labels are available, a final
optional stage freezes both the encoder and the task head and fits only the
residual \(r\). This staging reuses one scene representation across tasks
while confining expensive updates to the earliest pretraining stage.

\section{Experiments}
\subsection{Datasets and Protocols}
\begin{figure}[t]
\centering
\includegraphics[width=\linewidth]{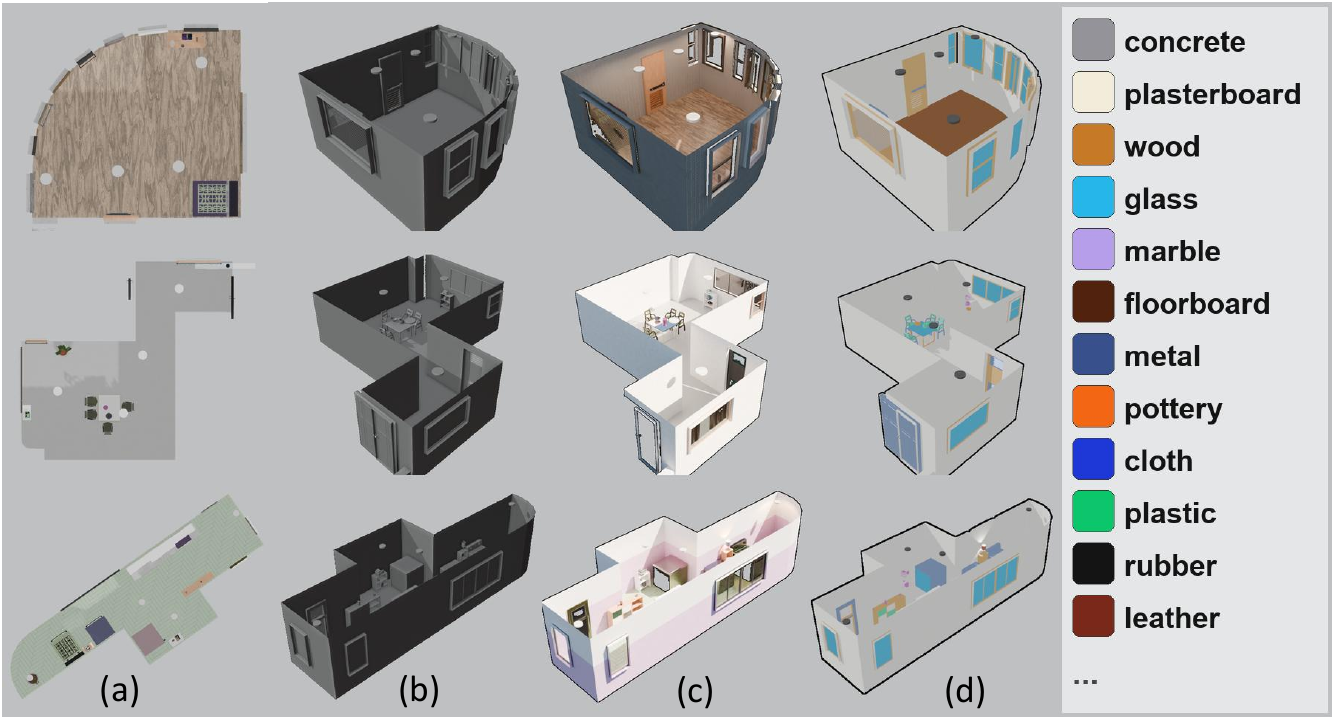}
\caption{PRISM dataset overview. Examples of: (a) floor plan;
(b) untextured meshes for a floor plan; (c) textured mesh objects;
(d) meshes colored by material class, with the legend on the right.}
\label{fig:dataset}
\end{figure}

We construct \textbf{PRISM} to provide paired geometry and radio labels for
cross-scene learning. Using Infinigen~\cite{infinigen2024indoors}, we generate
391 indoor rooms with mesh geometry, surface materials, and textures
(longest horizontal extents \(3\)--\(16\)\,m). Surface names map to 15
material classes with electromagnetic parameters from ITU-R
P.2040~\cite{itu-p2040} (Figure~\ref{fig:dataset}). Scenes are exported to
XML and labeled in Sionna RT~\cite{hoydis2023sionna} for PG and PAS.

\paragraph{PRISM-PG.}
For PG, Sionna's radio-map solver produces a dense \(16^{3}\) volume for each
TX at 3.5\,GHz, a representative mid-band 5G carrier, with up to 256 TX
locations per room. After filtering invalid simulations we retain 337 rooms
(269 / 33 / 35 train / val / test), totaling 86{,}272 TX-conditioned fields.
Point clouds sampled from surfaces serve as model inputs; meshes are used
only for offline labeling.

\paragraph{PRISM-PAS.}
For PAS, we sample \(20\times 20\) TX--RX pairs per room, compute channel
responses with Sionna's path solver, and form power spectra by conventional
beamforming on a \(4\times 4\) RX array (\(360\times 90\)
azimuth--elevation). Intersection with PRISM-PG yields 199 rooms that share
mesh geometry but use independently sampled TX--RX grids. A denser held-out
indoor room provides 2000 TX locations and 20 RXs for single-scene
comparison.

\subsection{Implementation and Baselines}

\paragraph{Model and training.}
The reported PG checkpoint uses width \(d{=}256\), CSC depth \(N{=}4\),
16 heads, FPS \(M{=}20{,}000\), downsampling rate 3, decoder neighborhood
\(K{=}8\), \(\log_{10}\sigma\) material channels, and TX--query geometry
(\(\approx\)29.8\,M parameters). PAS reuses the same backbone with a frozen
encoder and Fourier direction encoding of \(L{=}6\) frequencies. We optimize
with Adam, cosine learning-rate schedule, initial learning rate
\(2\times 10^{-4}\), MSE loss on normalized PG targets, batch size 16 with
data parallelism, and up to 3000 epochs; the reported checkpoint is selected
by best validation MAE. Scene splits and PAS train/test TX draws use
seed 2026. During development we mainly compared widths \(\{128,256\}\),
CSC depths \(\{2,4\}\), and neighborhood sizes \(K\in\{8,16\}\), and kept
the setting with the best validation MAE.
Unless noted, training and evaluation use PyTorch~2.4 on NVIDIA RTX~3090
GPUs; latency and peak-memory numbers in
Table~\ref{tab:inference_cost} are measured on a single RTX~3090
(batch size~1, warm steady state).

\paragraph{Baselines.}
We compare against learning methods rather than scoring a ray tracer on its
own labels. For cross-scene PG, we use two feedforward baselines trained on
the same scene-disjoint split: a RadioUNet-style slice-wise 2D
U-Net~\cite{RadioUNet} and a 3D U-Net on occupancy/material volumes. For
per-scene comparisons, we train NeRF\(^{2}\)~\cite{zhao2023nerf2},
WRF-GS+~\cite{11258087}, and GSRF~\cite{yang2026gsrf} on labeled links inside a target
room and evaluate held-out TXs. NeRF\(^{2}\) and WRF-GS+ appear in
both the PG sample-efficiency study and the PAS transfer study; GSRF is used
for PAS.

\paragraph{Metrics.}
We evaluate PG in dB after clipping predictions \(\hat{y}\) and labels \(y\)
to \([-90,0]\,\mathrm{dB}\). Over \(N\) evaluated samples, MAE
\begin{equation}
\mathrm{MAE}=\frac{1}{N}\sum_{i=1}^{N}|\hat{y}_{i}-y_{i}|
\end{equation}
is our primary metric, since radio coverage is conventionally reported on a
logarithmic power scale. We also report normalized mean squared error (NMSE)
and PSNR,
\begin{equation}
\mathrm{PSNR}=10\log_{10}\frac{R^{2}}{\frac{1}{N}\sum_{i=1}^{N}(\hat{y}_{i}-y_{i})^{2}},
\end{equation}
with dynamic range \(R{=}90\,\mathrm{dB}\) matching the clipped span. NMSE
removes absolute-scale sensitivity, while PSNR converts the same error into
a familiar logarithmic score. SSIM is computed on
horizontal \(16\times 16\) slices of each \(16^{3}\) volume with the same
range, to capture spatial structure beyond pointwise error. For PAS,
predictions and labels are scored in a per-pair normalized spectrum space on
\([0,1]\). We use SSIM under unit dynamic range for angular-lobe structure
and MAE for intensity error on the same maps. Tables report means over RXs,
and CDFs provide median and percentile behavior beyond a single average.

\subsection{PG Prediction}

\paragraph{Cross-scene feedforward baselines.}
We first compare against methods that, like Point2Radio, aim to generalize to
unseen scenes. Table~\ref{tab:main-pg} reports full-volume PG on the 35
held-out test rooms. Point2Radio\ reaches
0.871\,dB MAE and 0.954 SSIM (NMSE 0.0025, PSNR
34.83\,dB), reducing MAE by 76.7\% relative to RadioUNet
(3.745\,dB / 0.836) and remaining substantially more accurate than 3D~U-Net
(3.038\,dB / 0.884). Pure 2D slice models cannot resolve the underlying 3D
geometry, so even with height labels they struggle to capture vertical
structure and multipath. A volumetric 3D~U-Net provides a 3D representation,
yet still learns a largely scene-agnostic input--output map rather than
propagation structure, and therefore transfers poorly across rooms.
Figure~\ref{fig:vis} visualizes the same gap on three held-out rooms. 

\begin{table}[t]
\centering
{
\begin{tabular}{lcccc}
\toprule
Method & MAE$\downarrow$ & NMSE & PSNR$\uparrow$ & SSIM$\uparrow$ \\
\midrule
RadioUNet (2D) & 3.745 & 0.0159 & 25.10 & 0.836 \\
3D U-Net & 3.038 & 0.0152 & 26.71 & 0.884 \\
Point2Radio\ (ZS) & \textbf{0.871} & \textbf{0.0025} & \textbf{34.83} & \textbf{0.954}  \\
\bottomrule
\end{tabular}}
\caption{ZS full-volume PG prediction on the 35 held-out test scenes }
\label{tab:main-pg}
\end{table}

\begin{figure}[t]
    \centering
    \includegraphics[width=0.85\linewidth]{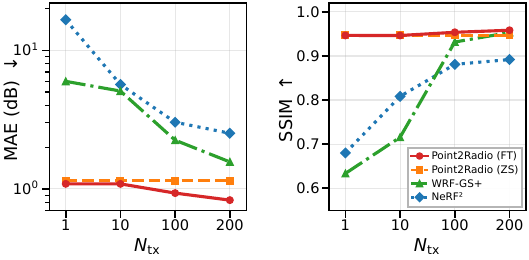}
    \caption{PG prediction accuracy on three held-out rooms under different
numbers of training TXs.}
    \label{fig:samples}
\end{figure}

\paragraph{Per-scene neural fields and sample efficiency.}
We next compare against per-scene methods that do not transfer across rooms.
NeRF\(^{2}\) and WRF-GS+ overfit a single environment from dense in-scene
supervision and then reconstruct radio fields for arbitrary TXs in that room.
Against this paradigm we report two operating modes for Point2Radio:
ZS cross-scene inference, and light target-scene adaptation with a
frozen backbone. To measure sample efficiency, we evaluate on three held-out
rooms with 256 TXs each: a fixed set of 56 TXs is reserved for testing, and
from the remaining 200 TXs we draw nested training sets of size
\(N_{\mathrm{tx}}\in\{1,10,100,200\}\) for per-scene training
(Figure~\ref{fig:samples}). Point2Radio\ already provides strong ZS
PG (\(\approx\)1.15\,dB MAE / 0.947 SSIM). As \(N_{\mathrm{tx}}\) grows,
NeRF\(^{2}\) and WRF-GS+ improve steadily; at \(N_{\mathrm{tx}}{=}200\),
WRF-GS+ exceeds our ZS SSIM (0.953 vs.\ 0.947) while still trailing in
MAE (1.57\,dB vs.\ 1.15\,dB). Fitting a lightweight residual head on the same
training TXs further improves Point2Radio, reducing MAE to 0.83\,dB and raising
SSIM to 0.958 at \(N_{\mathrm{tx}}{=}200\)
(\(\Delta\)MAE \(\approx\)0.32\,dB, \(\Delta\)SSIM \(\approx\)0.012 over
ZS).

\paragraph{Input ablation.} In practice it is difficult to obtain point clouds that jointly provide
accurate normals and electromagnetic material labels; materials are especially
hard to infer, even with vision priors and light calibration
\cite{an2026taming}. Table~\ref{tab:ablation} therefore removes ITU material
channels and/or normals at test time on the same cross-scene PG split,
keeping the trained weights fixed. Dropping materials raises MAE from
0.871\,dB to 1.259\,dB, dropping normals to 1.456\,dB, and removing both
to 1.685\,dB; SSIM falls from 0.954 to 0.943 / 0.933 / 0.920.

Normals thus matter more than material channels, consistent with their role
in reflection and occlusion, while materials still help for strong
reflectors such as metal. Even without either cue, Point2Radio\ remains
well below the U-Net baselines in Table~\ref{tab:main-pg}, showing that
positions alone carry a usable geometric prior and that normals and ITU
attributes act as refinements rather than hard prerequisites.

\begin{table}[t]
\centering
{
\begin{tabular}{lcc}
\toprule
Variant & MAE (dB)$\downarrow$ & SSIM$\uparrow$  \\
\midrule
Full Model & \textbf{0.871} & \textbf{0.954}  \\
w/o ITU materials & 1.259 & 0.943  \\
w/o normals & 1.456 & 0.933 \\
w/o materials \& normals & 1.685 & 0.920  \\
\bottomrule
\end{tabular}
}
\caption{Ablation of test-time inputs on cross-scene PG.
\label{tab:ablation}}
\end{table}

\subsection{PAS Prediction}

In PAS baselines, the RX is typically fixed while the TX moves,
and models are fit with dense in-scene supervision to reconstruct the spectrum
at that RX. We follow the same protocol in one held-out indoor room that is
unseen during Point2Radio\ pretraining. The room provides four RXs and
2000 TX locations per RX; for each RX we use an 80/20 TX split
(1600 train / 400 test) and report the mean over the four RXs.
Per-scene NeRF\(^{2}\), WRF-GS+, and GSRF are trained on the training TXs of
each RX. For Point2Radio\ we evaluate two modes with a frozen PG-pretrained
encoder: ZS inference, and light residual-head fine-tuning on the
training TXs.
Table~\ref{tab:pas} reports mean test SSIM and MAE over the four RXs.

\begin{table}[t]
\centering
{
\begin{tabular}{lcc}
\toprule
Method & SSIM$\uparrow$ & MAE$\downarrow$ \\
\midrule
NeRF$^{2}$ & 0.6775 & 0.0993 \\
WRF-GS+ & 0.7145 & 0.0872 \\
GSRF & 0.5686 & 0.1339 \\
Point2Radio\ (ZS) & 0.6899 & 0.1128 \\
Point2Radio\ (residual FT) & \textbf{0.7983} & \textbf{0.0798} \\
\bottomrule
\end{tabular}}
\caption{PAS prediction on one held-out indoor room (4 RXs; per-RX 80/20
train/test TX split). Mean over RXs.}
\label{tab:pas}
\end{table}

\begin{figure}[t]
\centering
\includegraphics[width=0.85\columnwidth]{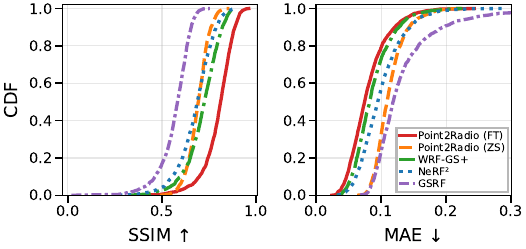}
\caption{CDF of per-pair test SSIM / MAE on one held-out indoor room
(4 RXs).}
\label{fig:aoa_cdf}
\end{figure}

\begin{figure}[t]
\centering
\includegraphics[width=\linewidth]{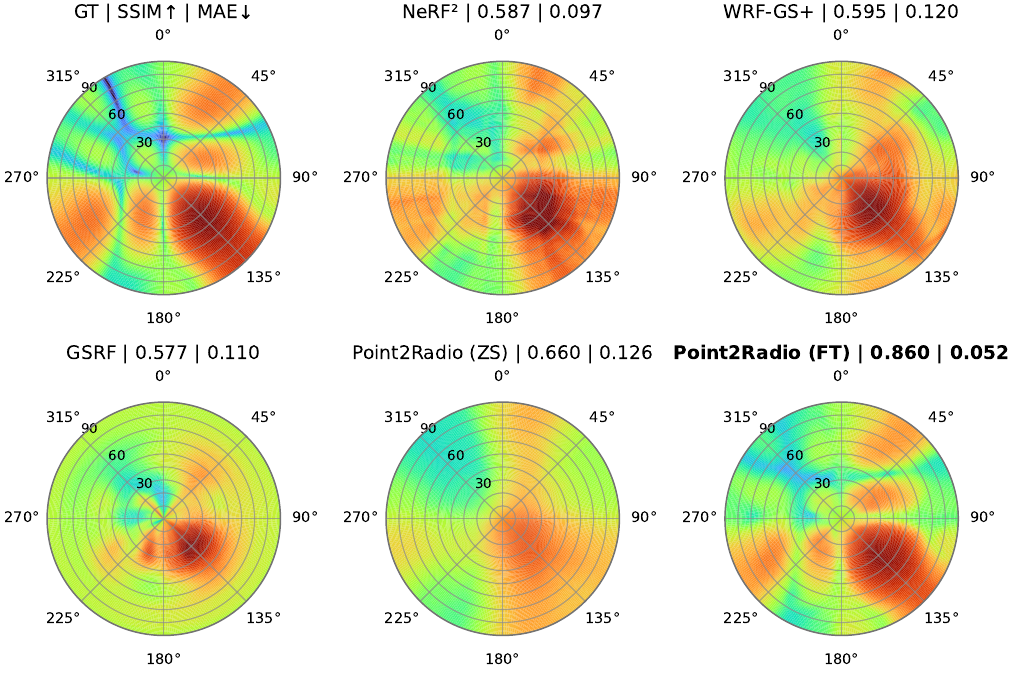}
\caption{PAS qualitative examples on held-out TX--RX pairs in one indoor room.}
\label{fig:aoa_vis}
\end{figure}

Figure~\ref{fig:aoa_vis} shows qualitative examples, and
Figure~\ref{fig:aoa_cdf} reports per-pair CDFs over the 1600 test pairs.
ZS Point2Radio\ reaches mean SSIM 0.6899 (median 0.6925),
slightly above NeRF\(^{2}\) (0.6775 / 0.6870) and well above GSRF
(0.5686 / 0.5821), while remaining close to WRF-GS+ (0.7145 / 0.7259). After
residual fine-tuning, Point2Radio\ rises to mean SSIM 0.7983
(median 0.8105, 90th 0.8866) and mean MAE 0.0798
(median 0.0745), exceeding WRF-GS+ by \(+\)0.084 mean SSIM and improving the
90th-percentile SSIM from 0.8206 to 0.8866. Relative to NeRF\(^{2}\) and GSRF
the mean-SSIM gains are \(+\)17.8\% and \(+\)40.4\%. The spatial backbone stays
frozen, so the gain comes from reusing the PG prior with a lightweight
scene-specific correction.

Together, the PG and PAS results support a consistent picture of what the
shared encoder has learned. Cross-scene PG accuracy shows that the frozen
backbone already captures how a TX interacts with 3D geometry and
materials, rather than memorizing a single room. Reusing that encoder for PAS
with only a lightweight task head then yields competitive ZS spectra
and strong gains after residual adaptation, indicating that the same
TX--scene interaction prior transfers across radio quantities. Overall,
Point2Radio\ stays compact in parameters and inference cost, yet delivers
high accuracy across tasks and scenes through a reusable propagation
representation plus light task- and scene-specific heads.

\subsection{Inference Efficiency}

Table~\ref{tab:inference_cost} compares feedforward Point2Radio\ with the
Sionna~RT labeling pipelines on an RTX~3090, batch size~1 in warm steady
state. For one TX-conditioned \(16^{3}\) PG volume, Point2Radio\ finishes in
163\,ms at 3.3\,GB peak memory, versus 990\,ms / 8.3\,GB for Sionna, about
\(6.1\times\) faster and \(2.5\times\) less memory. For one PAS pair the gap
is 156\,ms / 1.2\,GB versus 648\,ms / 3.3\,GB, about \(4.2\times\) faster.
Inference uses only a point cloud and transceiver queries, with no mesh and
no online path tracing.

The speedup comes from where the compute sits. Encoding a TX-conditioned
scene representation takes about 151\,ms, while decoding a full \(16^{3}\)
volume from the cached codes takes only about 10\,ms, and a single RX query
about 3\,ms. Because most of the work is in the encoder and the local
decoder is cheap, dense querying under a fixed TX does not grow linearly
with the number of RXs. The model encodes once and then evaluates
many locations quickly on the decoder side.

Combined with the PG accuracy above, this makes Point2Radio\ attractive as a
fast surrogate for large-scale radio-field generation. Regenerating the
86{,}272 TX-conditioned volumes in PRISM-PG at these steady-state rates would
take about 24\,hours with Sionna~RT, but only about 4\,hours with
Point2Radio, a roughly \(6\times\) reduction in wall-clock labeling time
while retaining high fidelity on held-out scenes.

\begin{table}[t]
\centering
{
\begin{tabular}{@{}lcccc@{}}
\toprule
& \multicolumn{2}{c}{PG ($16^{3}$)} & \multicolumn{2}{c}{PAS (1 pair)} \\
\cmidrule(lr){2-3}\cmidrule(lr){4-5}
Method & Time & Peak GPU & Time & Peak GPU \\
\midrule
Our encode & 151\,ms & --- & 150\,ms & --- \\
Our decode & 10\,ms & --- & 3\,ms & --- \\
Our end-to-end & 163\,ms & 3.3\,GB & 156\,ms & 1.2\,GB \\
Sionna RT & 990\,ms & 8.3\,GB & 648\,ms & 3.3\,GB \\
\bottomrule
\end{tabular}
}
\caption{Inference latency and peak GPU memory on an RTX~3090.}
\label{tab:inference_cost}
\end{table}

\section{Conclusion and Future Work}

We present Point2Radio, a foundation model that learns transferable
TX-conditioned radio fields from material-aware point clouds. Hierarchical
tokenization and CSC attention produce a reusable scene
representation, which task-specific query heads map to dense PG
fields and PAS, with light residual adaptation when target
labels are available. This suggests that, with sufficiently diverse data, a
model can capture accurate propagation structure without explicitly
executing deterministic physics at inference. Our experiments show strong
cross-scene accuracy and efficiency. Point2Radio\ achieves 0.871\,dB MAE
and 0.954 SSIM for PG, reducing error by 76.7\% relative to a UNet-style
baseline, transfers effectively to PAS, and runs several times faster than
Sionna~RT at lower peak memory, without meshes or online path tracing.
Future work will focus on the following directions.
\begin{itemize}
    \item \textbf{Input modality.}
    Our pipeline currently assumes a prepared point cloud, which is easier
    than a simulation mesh but still nontrivial in practical deployments.
    Future work may take more accessible modalities such as images and
    videos as input, enabling Point2Radio\ to support a broader range of
    downstream tasks.
    \item \textbf{Training and adaptation data.}
    We aim to train on larger, higher-quality multimodal corpora to reduce
    dependence on simulator fidelity, and to use real radio measurements
    for rapid fine-tuning across diverse downstream tasks.
\end{itemize}

\bibliographystyle{plainnat}
\bibliography{paper}

\end{document}